\documentclass[pre,amsmath,amssymb,reprint]{revtex4-2}
\usepackage{epsfig}                                                                 
\usepackage{dcolumn}                                                                
\usepackage{bm}                                                                     
\let\boldsymbol\pmb
\usepackage[pdfstartview=FitH,bookmarksopen=true,bookmarksopenlevel=1]{hyperref}    
\begin{document}
\title[Dynamics of a drop floating in vapor of the same fluid]{Dynamics of a drop floating in vapor of the same fluid}
\author{E.~S.~Benilov}
 \email[Email address: ]{Eugene.Benilov@ul.ie}
 \homepage[\newline Homepage: ]{https://staff.ul.ie/eugenebenilov/}
 \affiliation{Department of Mathematics and Statistics, University of Limerick, Limerick V94~T9PX, Ireland}

\date{\today}

\begin{abstract}
Evaporation of a liquid drop surrounded by either vapor of the same fluid, or
vapor and air, is usually attributed to vapor diffusion -- which, however,
does not apply to the former setting, as pure fluids do not diffuse. The
present paper puts forward an additional mechanism, one that applies to both
settings. It is shown that disparities between the drop and vapor in terms of
their pressure and chemical potential give rise to a flow. Its direction
depends on the vapor density and the drop's size. In undersaturated or
saturated vapor, all drops evaporate -- but in oversaturated (yet
thermodynamically stable) vapor, there exists a critical radius: smaller drops
evaporate, larger drops act as centers of condensation and grow. The developed
model is used to estimate the evaporation time of a drop floating in saturated
vapor. It is shown that, if the vapor-to-liquid density ratio is small, so is
the evaporative flux -- as a result, millimeter-sized water drops at
temperatures lower than $70^{\circ}\mathrm{C}$ survive for days. If, however,
the temperature is comparable (but not necessarily close) to its critical
value, such drops evaporate within minutes. Micron-sized drops, in turn,
evaporate within seconds for all temperatures between the triple and critical points.

\end{abstract}
\maketitle

\section{Introduction}

It is well known that liquid drops surrounded by under-saturated vapor or air
evaporate. If, on the other hand, the vapor (or air) is saturated, it is
natural to assume that it would be in equilibrium with the liquid, and no
evaporation would occur. Yet it does -- due to the so-called Kelvin effect,
shifting the saturation conditions due to the capillary pressure associated
with the curvature of the drop's boundary
\cite{EggersPismen10,ColinetRednikov11,RednikovColinet13,Morris14,JanecekDoumencGuerrierNikolayev15,RednikovColinet17,RednikovColinet19}%
.

To understand the underlying physics, compare a flat liquid/vapor interface
with a spherical one. Let the temperature $T$ in both cases be spatially uniform.

The flat interface is in \emph{mechanical} equilibrium if the liquid and vapor
pressures are in balance,%
\begin{equation}
p(\rho_{l.sat})=p(\rho_{v.sat}), \label{1.1}%
\end{equation}
where $\rho_{_{v.sat}}$ is the density of saturated vapor, $\rho_{l.sat}$ is
that of liquid, and $p$ is implied to also depend on $T$.

The \emph{thermodynamic} equilibrium, in turn, requires isothermality (already
assumed), plus the equality of the specific chemical potentials,%
\begin{equation}
G(\rho_{l.sat})=G(\rho_{v.sat}).\label{1.2}%
\end{equation}
Physically, $G$ represents the fluid's energy per unit mass, so that condition
(\ref{1.2}) guarantees that a molecule can be moved through the interface
without changing the system's net energy. Conditions (\ref{1.1})--(\ref{1.2})
have been formulated by Maxwell \cite{Maxwell75}, and are often referred to as
the \textquotedblleft Maxwell construction\textquotedblright.

Next, consider a \emph{spherical} liquid drop floating in saturated vapor.
Condition (\ref{1.1}) in this case should be replaced with%
\begin{equation}
p(\rho_{l})=p(\rho_{v})+\frac{2\sigma}{r_{d}}, \label{1.3}%
\end{equation}
where $\sigma$ is the fluid's surface tension, $r_{d}$ is the drop's radius,
and the subscript $_{.sat}$ was omitted because the `new' densities of liquid
and vapor do not have to coincide with their saturated values. Condition
(\ref{1.2}), in turn, does not depend on the shape of the interface and, thus,
applies to spherical drops as well,
\begin{equation}
G(\rho_{l})=G(\rho_{v}). \label{1.4}%
\end{equation}
Conditions (\ref{1.3})--(\ref{1.4}) form a `new' version of the Maxwell construction.

Next, imagine that the drop is surrounded by \emph{saturated} vapor ($\rho
_{v}=\rho_{v.sat}$), in which case the only remaining unknown, $\rho_{l}$,
cannot satisfy both equations (\ref{1.3})--(\ref{1.4}) for any value of the
drop's radius except $r_{d}=\infty$. Thus, a volume of liquid with a curved
boundary cannot be in both mechanical and thermodynamic equilibria with
saturated vapor. As a result, a flow develops, giving rise to a
liquid-to-vapor (evaporative) flux.

In principle, this flux can be `smothered' by increasing the pressure of the
surrounding vapor beyond the saturation level (provided it is still
subspinodal, i.e., thermodynamically stable). In this case, a finite $r_{d}$
may exist such that both conditions (\ref{1.3})--(\ref{1.4}) hold, and so the
evaporative flux is zero. It still remains unclear what happens with drops
with radii different from this `special' value of $r_{d}$. Do larger drops
recede and smaller ones grow (thus, making the special value of $r_{d}$ an
attractor) -- or is it the other way around?

This question is answered in the present paper.

In order to place its findings in the context of the existing literature, note
that numerous authors examined evaporation of \emph{sessile} drops surrounded
by undersaturated or saturated air, e.g., Refs.
\cite{BurelbachBankoffDavis88,DeeganBakajinDupontHuberEtal00,Ajaev05,DunnWilsonDuffyDavidSefiane09,EggersPismen10,CazabatGuena10,ColinetRednikov11,MurisicKondic11,RednikovColinet13,Morris14,StauberWilsonDuffySefiane14,StauberWilsonDuffySefiane15,JanecekDoumencGuerrierNikolayev15,SaxtonWhiteleyVellaOliver16,SaxtonVellaWhiteleyOliver17,BrabcovaMchaleWellsBrown17,RednikovColinet19,WrayDuffyWilson19}%
. These studies are based on the lubrication approximation (applicable to
\emph{thin} sessile drops), complemented with the assumption that the vapor
density just above the drop's surface exceeds $\rho_{v.sat}$ by a certain
amount depending on the interfacial curvature. This value is then used as a
boundary condition for the diffusion equation describing the surrounding vapor.

Alternatively -- as done in the present paper -- one can use the
diffuse-interface model (DIM). It assumes that the fluid density varies
smoothly in the interfacial region, thus providing an integral description of
vapor, liquid, and the mass exchange between them.

Physically, the DIM is based on Korteweg's assumption \cite{Korteweg01} that
the van der Waals (intermolecular) force can be approximated by a pair-wise
potential whose spatial scale is much smaller than the interfacial thickness.
It had been later used in a large number of important applications: phase
transitions of the second kind \cite{Ginzburg60}, spinodal decomposition
\cite{Cahn61}, nucleation and collapse of bubbles
\cite{MagalettiMarinoCasciola15,MagalettiGalloMarinoCasciola16,GalloMagalettiCasciola18,GalloMagalettiCoccoCasciola20,GalloMagalettiCasciola21}%
, phase separation of polymers \cite{ThieleMadrugaFrastia07,MadrugaThiele09},
contact lines
\cite{AndersonMcFaddenWheeler98,PismenPomeau00,DingSpelt07,YueZhouFeng10,YueFeng11,SibleyNoldSavvaKalliadasis14,BorciaBorciaBestehornVarlamovaHoefnerReif19}%
, Faraday instability
\cite{BorciaBestehorn14,BestehornSharmaBorciaAmiroudine21}, etc. Recently the
DIM was shown to follow from the kinetic theory of dense gases
\cite{Giovangigli20,Giovangigli21} via the standard Chapman--Enskog asymptotic method.

The DIM has also been used for settings involving the Kelvin effect: to prove
the nonexistence (due to evaporation) of steady solutions for two- and
three-dimensional sessile drops surrounded by saturated vapor
\cite{Benilov20c,Benilov21b}, to derive an asymptotic description of liquid
films \cite{Benilov20d,Benilov22b}, and to show that vapor can spontaneously
condensate in a corner formed by two walls \cite{Benilov22b}. The DIM allows
one to create a comprehensive model from scratch -- instead of building it
from `blocks' describing liquid, vapor, their interactions with one another
and the substrate, etc.

The present paper has the following structure: in Sec. \ref{Sec 2}, the
problem will be formulated mathematically. Sec. \ref{Sec 3} examines
analytically and numerically static drops floating in oversaturated vapor. In
Sec. \ref{Sec 4}, the dynamics of evolving drops is explored asymptotically
under the assumption that the drop's radius exceeds the interfacial thickness,
and the vapor density is close, but not necessarily equal, to its saturated
value. Sec. \ref{Sec 5} presents an estimate of evaporation times of real
water drops.

\section{Formulation\label{Sec 2}}

\subsection{The governing equations}

Thermodynamically, the state of a compressible non-ideal fluid is fully
described by its density $\rho$ and temperature $T$. The dependence of the
pressure $p$ on $\left(  \rho,T\right)  $, or the equation of state, is
assumed to be known -- as are those of the shear viscosity $\mu_{s}$ and the
bulk viscosity $\mu_{b}$.

Consider a flow characterized by its density field $\rho(\mathbf{r},t)$ and
velocity field $\mathbf{v}(\mathbf{r},t)$, where $\mathbf{r}=\left[
x,y,z\right]  $ is the coordinate vector and $t$, the time. Let the flow be
affected by a force $\mathbf{F}$ (to be later identified with the van der
Waals force). Under the assumption that the temperature is uniform in both
space and time, the flow is governed by the following equations:%
\begin{equation}
\frac{\partial\rho}{\partial t}+\boldsymbol{\boldsymbol{\nabla}}\cdot\left(
\rho\mathbf{v}\right)  =0, \label{2.1}%
\end{equation}%
\begin{equation}
\frac{\partial\mathbf{v}}{\partial t}+\left(  \mathbf{v}\cdot
\boldsymbol{\boldsymbol{\nabla}}\right)  \mathbf{v}+\frac{1}{\rho
}\boldsymbol{\boldsymbol{\nabla}}p=\frac{1}{\rho}%
\boldsymbol{\boldsymbol{\nabla}}\cdot\boldsymbol{\Pi}+\mathbf{F}, \label{2.2}%
\end{equation}
where the viscous stress tensor is%
\begin{equation}
\boldsymbol{\Pi}=\mu_{s}\left[  \boldsymbol{\boldsymbol{\nabla}}%
\mathbf{v}+\left(  \boldsymbol{\boldsymbol{\nabla}}\mathbf{v}\right)
^{T}-\frac{2}{3}\mathbf{I}\left(  \boldsymbol{\boldsymbol{\nabla}}%
\cdot\mathbf{v}\right)  \right]  +\mu_{b}\,\mathbf{I}\left(
\boldsymbol{\boldsymbol{\nabla}}\cdot\mathbf{v}\right)  , \label{2.3}%
\end{equation}
and $\mathbf{I}$ is the identity matrix. Within the framework of the DIM, the
van der Waals force is given by%
\begin{equation}
\mathbf{F}=K\boldsymbol{\boldsymbol{\nabla}}\nabla^{2}\rho, \label{2.4}%
\end{equation}
where the Korteweg parameter $K$ is determined by the fluid's molecular
structure and is related to (can be calculated from) the surface tension and
equation of state. In the DIM, $\mathbf{F}$ is responsible for phase
transitions, interfacial dynamics, and all related phenomena including the
Kelvin effect.

Set (\ref{2.1})--(\ref{2.4}) has been first proposed by Pismen and Pomeau
\cite{PismenPomeau00}, and can be viewed as the isothermal reduction of the
set of equations \cite{Desobrino76}. The condition when the DIM can be assumed
isothermal has been derived in Ref. \cite{Benilov20b}: it implies that the
flow is sufficiently slow, so that, first, viscosity does not produce much
heat, and second, heat conduction has enough time to homogenize the
temperature field.

To nondimensionalize the problem, introduce a characteristic density $\varrho$
and pressure $P$. Together with the Korteweg parameter, they define a spatial
scale%
\[
l=\frac{K^{1/2}\varrho}{P^{1/2}},
\]
representing characteristic thickness of a liquid/vapor interface. Estimates
\cite{MagalettiGalloMarinoCasciola16,GalloMagalettiCoccoCasciola20,Benilov20a}
show that $l$ is on a nanometer scale.

The following nondimensional variables will be used:%
\begin{gather*}
r_{nd}=\frac{r}{l},\qquad t_{nd}=\frac{V}{L}t,\\
\rho_{nd}=\frac{\rho}{\varrho},\qquad T_{nd}=\frac{\varrho RT}{P},\\
p_{nd}=\frac{p}{P},\qquad\mathbf{v}_{nd}=\frac{\mathbf{v}}{V},\qquad
\mathbf{\Pi}_{nd}=\frac{L^{2}\mathbf{\Pi}}{K\varrho},
\end{gather*}
where $L$ is the characteristic drop radius, $R$ is the specific gas constant,
and the velocity scale is%
\[
V=\frac{\left(  K\varrho\right)  ^{1/2}}{L}.
\]
Introduce also nondimensional viscosities,%
\[
\left(  \mu_{s}\right)  _{nd}=\frac{\mu_{s}}{K^{1/2}\varrho^{3/2}}%
,\qquad\left(  \mu_{b}\right)  _{nd}=\frac{\mu_{b}}{K^{1/2}\varrho^{3/2}}.
\]
Rewriting equations (\ref{2.1})--(\ref{2.4}) in terms of the nondimensional
variables and omitting the subscript $_{nd}$, one obtains%
\begin{equation}
\varepsilon\frac{\partial\rho}{\partial t}+\boldsymbol{\boldsymbol{\nabla}%
}\cdot\left(  \rho\mathbf{v}\right)  =0, \label{2.5}%
\end{equation}
\begin{widetext}%
\begin{equation}
\varepsilon^{2}\left[ \varepsilon \frac{\partial\mathbf{v}}{\partial t}+\left(
\mathbf{v}\cdot\boldsymbol{\boldsymbol{\nabla}}\right)  \mathbf{v}\right]
+\frac{1}{\rho}\boldsymbol{\boldsymbol{\nabla}}p=\frac{\varepsilon}{\rho
}\boldsymbol{\boldsymbol{\nabla}}\cdot\left\{  \mu_{s}\left[
\boldsymbol{\boldsymbol{\nabla}}\mathbf{v}+\left(
\boldsymbol{\boldsymbol{\nabla}}\mathbf{v}\right)  ^{T}-\frac{2}{3}%
\mathbf{I}\left(  \boldsymbol{\boldsymbol{\nabla}}\cdot\mathbf{v}\right)
\right]  +\mu_{b}\,\mathbf{I}\left(  \boldsymbol{\boldsymbol{\nabla}}%
\cdot\mathbf{v}\right)  \right\}  +\boldsymbol{\boldsymbol{\nabla}}\nabla
^{2}\rho,\label{2.6}%
\end{equation}
\end{widetext}where%
\[
\varepsilon=\frac{l}{L}.
\]
In what follows, it is convenient to introduce the chemical potential
$G(\rho)$ related to the pressure by%
\begin{equation}
\frac{\partial G}{\partial\rho}=\frac{1}{\rho}\frac{\partial p}{\partial\rho}.
\label{2.7}%
\end{equation}
Given the equation of state $p(\rho)$, this equality fixes $G(\rho)$ up to an
arbitrary function of the temperature (which is unimportant as $T$ is assumed
to be constant).

In this paper, the general results will be illustrated using the
(nondimensional) van der Waals equation of state,%
\begin{equation}
p=\frac{T\rho}{1-\rho}-\rho^{2}, \label{2.8}%
\end{equation}
in which case the chemical potential can be chosen in the form%
\begin{equation}
G=T\left(  \ln\frac{\rho}{1-\rho}+\frac{1}{1-\rho}\right)  -2\rho. \label{2.9}%
\end{equation}

\subsection{Flat liquid/vapor interface}

In what follows, the solution will be needed describing a static flat
interface separating liquid and vapor of the same fluid, in an unbounded space.

Let the fluid be at rest ($\mathbf{v}=\mathbf{0}$) and its density field,
independent of $t$, $x$, and $y$. Letting accordingly%
\[
\rho=\bar{\rho}(z),
\]
one can reduce equations (\ref{2.5})--(\ref{2.6}) to%
\begin{equation}
\frac{1}{\bar{\rho}}\frac{\mathrm{d}p(\bar{\rho})}{\mathrm{d}z}=\frac
{\mathrm{d}^{3}\bar{\rho}}{\mathrm{d}z^{3}}. \label{2.10}%
\end{equation}
To single out the solution describing a liquid/vapor interface, the following
boundary conditions will be imposed:%
\begin{align}
\bar{\rho}  &  \rightarrow\rho_{l.sat}\qquad\text{as}\qquad z\rightarrow
-\infty,\label{2.11}\\
\bar{\rho}  &  \rightarrow\rho_{v.sat}\qquad\text{as}\qquad z\rightarrow
+\infty, \label{2.12}%
\end{align}
where $\rho_{v.sar}$ is the saturated vapor density and $\rho_{l.sat}$ is its
liquid counterpart. Boundary-value problem (\ref{2.10})--(\ref{2.12}) is
invariant with respect to the change $z\rightarrow z+\operatorname{const}$ --
hence, to uniquely fix $\bar{\rho}(z)$, one should impose an extra condition,
say,%
\begin{equation}
\bar{\rho}=\frac{1}{2}\left(  \rho_{v.sat}+\rho_{l.sat}\right)  \qquad
\text{at}\qquad z=0. \label{2.13}%
\end{equation}
Boundary-value problem (\ref{2.10})--(\ref{2.13}) is of fundamental
importance, as it gives rise to the Maxwell construction. This can be shown
using the first integrals of equation (\ref{2.10}), of which there are two. To
derive the first one, multiply (\ref{2.10}) by $\bar{\rho}$, integrate it with
respect to $z$, and fix the constant of integration via boundary condition
(\ref{2.12}). Eventually, one obtains%
\begin{equation}
p(\bar{\rho})-\bar{\rho}\frac{\mathrm{d}^{2}\bar{\rho}}{\mathrm{d}z^{2}}%
+\frac{1}{2}\left(  \frac{\mathrm{d}\bar{\rho}}{\mathrm{d}z}\right)
^{2}=p(\rho_{v.sat}). \label{2.14}%
\end{equation}
Another first integral can be derived by rearranging equation (\ref{2.10})
using identity (\ref{2.7}) and integrating. The constant of integration can
again be fixed via boundary condition (\ref{2.12}), and one obtains%
\begin{equation}
G(\bar{\rho})-\frac{\mathrm{d}^{2}\bar{\rho}}{\mathrm{d}z^{2}}=G(\rho
_{v.sat}). \label{2.15}%
\end{equation}
Considering equalities (\ref{2.14})--(\ref{2.15}) in the limit $z\rightarrow
-\infty$ and recalling boundary condition (\ref{2.11}), one obtains the
Maxwell construction (\ref{1.1})--(\ref{1.2}), as required.

To ensure that $\rho_{v.sat}$ and $\rho_{l.sat}$ are \emph{uniquely}
determined by equations (\ref{1.1})--(\ref{1.2}), one should also let
$\rho_{v.sat}<\rho_{l.sat}$ and require that both liquid and vapor be
thermodynamically stable. The latter condition implies%
\[
\left(  \frac{\partial p}{\partial\rho}\right)  _{\rho=\rho_{v.sat}}%
\geq0,\qquad\left(  \frac{\partial p}{\partial\rho}\right)  _{\rho
=\rho_{l.sat}}\geq0,
\]
i.e., the pressure may not decrease with density.

In what follows, the asymptotics of $\bar{\rho}(z)$ as $z\rightarrow-\infty$
will be needed. One can readily deduce from equation (\ref{2.15}) and boundary
conditions (\ref{2.11}) that\begin{widetext}%
\begin{equation}
\rho\rightarrow\rho_{l.sat}+A_{l}\exp\left[  \sqrt{\left(  \frac{\partial
G}{\partial\rho}\right)  _{\rho=\rho_{l.sat}}}z\right]  \qquad\text{as}\qquad
z\rightarrow-\infty,\label{2.16}%
\end{equation}%
\end{widetext}where $A_{l}$ and $A_{v}$ are undetermined constants. Note that
the expression under the square roots in asymptotics (\ref{2.16}) is positive
because equality (\ref{2.7}) makes $\partial G/\partial\rho$ have the same
sign as $\partial p/\partial\rho$, and the latter is positive at $\rho
=\rho_{v.sat}$ and $\rho=\rho_{l.sat}$ because saturated vapor and liquid are stable.

Finally, introduce the\emph{ }spinodal vapor density $\rho_{v.spi}$ as the
smaller root of the equation%
\[
\left(  \frac{\partial p}{\partial\rho}\right)  _{\rho=\rho_{v.spi}}=0,
\]
and the larger root corresponds to $\rho_{l.spi}$. Now, the stability
condition of vapor (any vapor, not just the saturated one) can be written in
the form $\rho_{v}\leq\rho_{v.spi}$, and a liquid is stable if $\rho_{l}%
\geq\rho_{l.spi}$.

\begin{figure}
\includegraphics[width=\columnwidth]{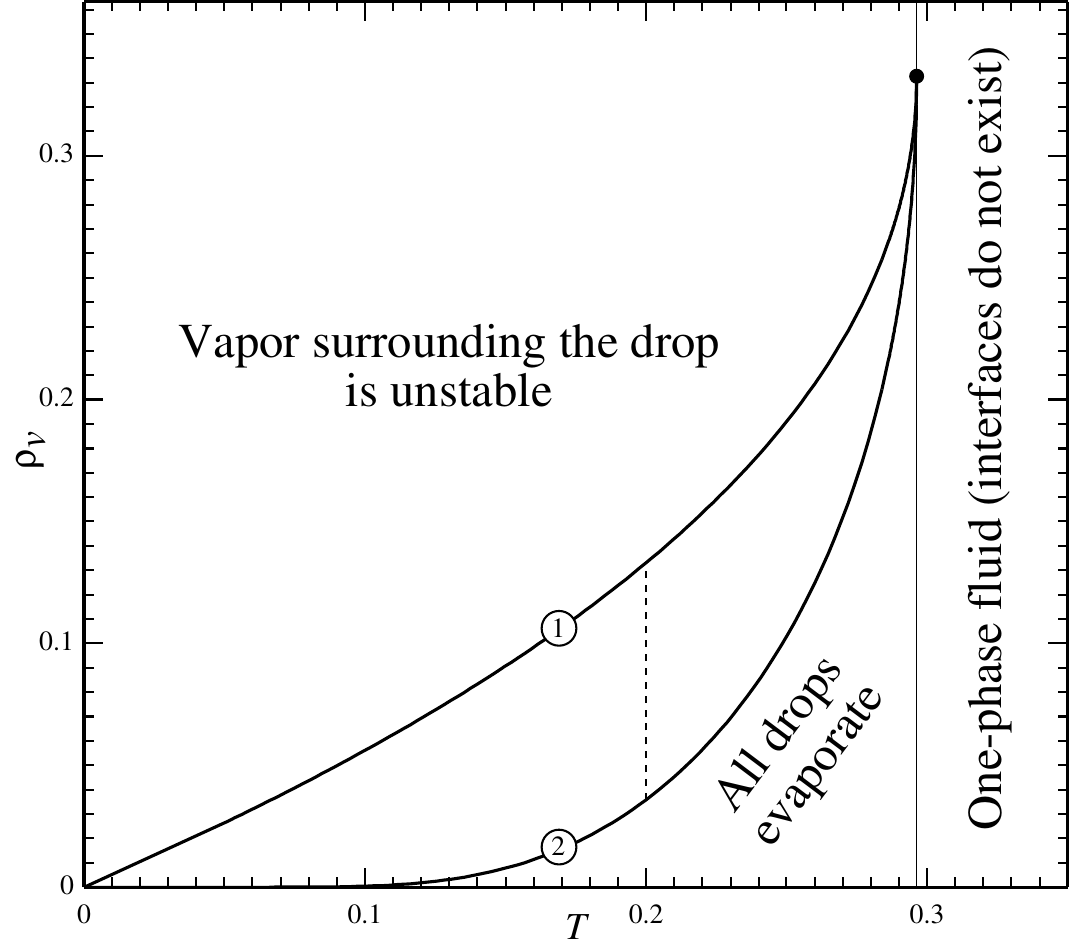}
\caption{The parameter space of the drop--vapor system, for the van der Waals fluid (\ref{2.8})--(\ref{2.9}). $T$ is the nondimensional temperature, $\rho_{v}$ is the nondimensional density of the vapor surrounding the drop. (1) the spinodal curve $\rho _{v}=\rho_{v.spi}$, (2) the saturation curve $\rho_{v}=\rho_{v.sat}$. The black dot marks the critical point. The dotted line marks the cross-section of the parameter space corresponding to Figs. \ref{fig2}--\ref{fig3}.}
\label{fig1}
\end{figure}

The saturated and spinodal densities of vapor have been computed for the van
der Waals fluid and plotted in Fig. \ref{fig1} on the $\left(  T,\rho
_{v}\right)  $ plane -- which also happens to be the parameter space of the
problem under consideration (drops floating in vapor).

Since drops in \emph{under}saturated vapor quickly evaporate and vapor with
$\rho_{v}>\rho_{v.spi}$ is unstable, the only region of the parameter space
where a non-trivial behavior can be observed is that between the two curves
depicted in Fig. \ref{fig1}.

\subsection{Spherically-symmetric flows}

Let the density field depend on $t$ and the radial variable $r=\left\vert
\mathbf{r}\right\vert $, and let the flow involve a single component of the
velocity $v=\left\vert \mathbf{v}\right\vert $, also depending on $r$ and $t$.
The corresponding reductions of the governing equations (\ref{2.5}%
)--(\ref{2.6}) are \cite{Heinbockel01}%
\begin{equation}
\varepsilon\frac{\partial\rho}{\partial t}+\frac{\partial\left(  v\rho\right)
}{\partial r}+\frac{2v\rho}{r}=0, \label{2.17}%
\end{equation}
\begin{widetext}%
\begin{multline}
\varepsilon^{2}\left(  \varepsilon\frac{\partial v}{\partial t}+v\frac
{\partial v}{\partial r}\right)  +\frac{1}{\rho}\frac{\partial p}{\partial
r}=\frac{\varepsilon}{\rho}\left\{  \frac{\partial}{\partial r}\left[
\frac{4\mu_{s}}{3}\left(  \frac{\partial v}{\partial r}-\frac{v}{r}\right)
+\mu_{b}\left(  \frac{\partial v}{\partial r}+\frac{2v}{r}\right)  \right]
+\frac{4\mu_{s}}{r}\left(  \frac{\partial v}{\partial r}-\frac{v}{r}\right)
\right\}  \\
+\frac{\partial}{\partial r}\left(  \frac{\partial^{2}\rho}{\partial r^{2}%
}+\frac{2}{r}\frac{\partial\rho}{\partial r}\right)  ,\label{2.18}%
\end{multline}
\end{widetext}Equations (\ref{2.17})--(\ref{2.18}) should be complemented by
the smoothness conditions at the origin,%
\begin{align}
\frac{\partial\rho}{\partial r}  &  =0\qquad\text{at}\qquad r=0,\label{2.19}\\
v  &  =0\qquad\text{at}\qquad r=0. \label{2.20}%
\end{align}
At infinity, the fluid's density becomes uniform and the viscous stress, zero,%
\begin{align}
\rho &  \rightarrow\rho_{v}\qquad\text{as}\qquad r\rightarrow\infty
,\label{2.21}\\
\frac{\partial v}{\partial r}  &  \rightarrow0\qquad~\,\text{as}\qquad
r\rightarrow\infty, \label{2.22}%
\end{align}
where $\rho_{v}$ is the vapor density.

Given suitable initial data, equations (\ref{2.17})--(\ref{2.18}) and boundary
conditions (\ref{2.19})--(\ref{2.22}) fully determine $\rho(r,t)$ and $v(r,t)$.

\section{Static drops\label{Sec 3}}

Static drops correspond to $v=0$ and $\partial\rho/\partial t=0$, in which
case equation (\ref{2.17}) holds automatically and (\ref{2.18}) reduces to%
\[
\frac{1}{\rho}\frac{\mathrm{d}p}{\mathrm{d}r}=\frac{\mathrm{d}}{\mathrm{d}%
r}\left(  \frac{\mathrm{d}^{2}\rho}{\mathrm{d}r^{2}}+\frac{2}{r}%
\frac{\mathrm{d}\rho}{\mathrm{d}r}\right)  .
\]
Rearrange this equation using definition (\ref{2.7}) of $G$, then integrate
and fix the constant of integration via condition (\ref{2.21}) -- which yields%
\begin{equation}
G(\rho)-G(\rho_{v})=\frac{\mathrm{d}^{2}\rho}{\mathrm{d}r^{2}}+\frac{2}%
{r}\frac{\mathrm{d}\rho}{\mathrm{d}r}. \label{3.1}%
\end{equation}
This ordinary differential equation and conditions (\ref{2.19}), (\ref{2.21})
form a boundary-value problem for $\rho(r)$.

As shown Appendix \ref{Appendix A}, problem (\ref{3.1}), (\ref{2.19}),
(\ref{2.21}) admits a nontrivial solution (such that $\rho(r)\neq\rho_{v}$ for
some $r$) only if%
\begin{equation}
\rho_{v.sat}<\rho_{v}<\rho_{v.spi}. \label{3.2}%
\end{equation}
To interpret this result physically, recall that a liquid drop cannot be
static in \emph{under}saturated vapor ($\rho_{v}<\rho_{v.sat}$), as it
evaporates. Drops in saturated vapor ($\rho_{v}=\rho_{v.sat}$) also evaporate
-- due to the Kelvin effect, as argued in this paper -- and this conclusion is
in line with similar nonexistence theorems for static drops on a substrate
\cite{Benilov20c,Benilov21b}. As for the range $\rho_{v}\geq\rho_{v.spi}$, the
vapor is unstable in this case, and so the drop should grow due to spinodal
decomposition. This leaves (\ref{3.2}) as the only possible range of $\rho
_{v}$ where drops can be static.

\begin{figure}
\includegraphics[width=\columnwidth]{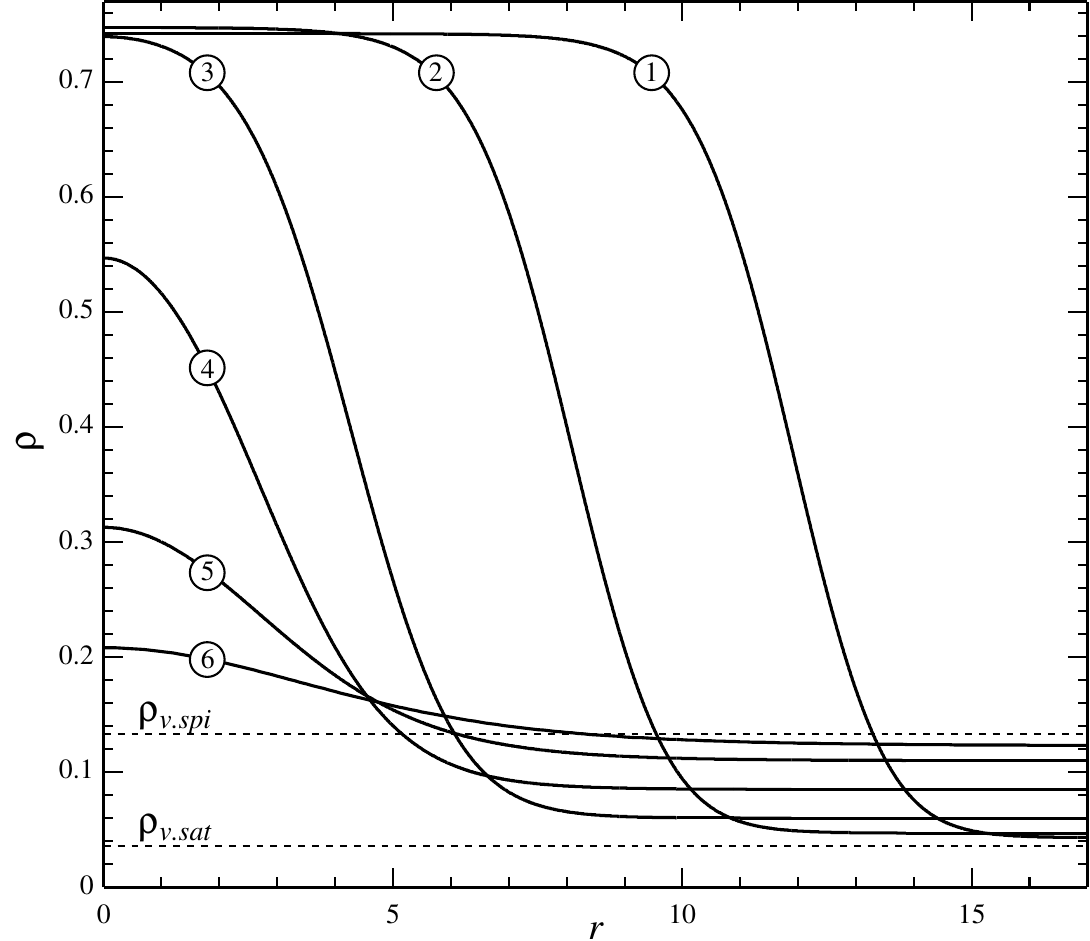}
\caption{Examples of drop profiles $\rho(r)$ for the van der Waals fluid at $T=0.2$. Curves (1)--(6) correspond to $\rho_{v}=0.043$, $0.470$, $0.6$, $0.85$, $0.11$, and $0.123$, respectively ($\rho_{v}$ is the density of the vapor surrounding the drop). The examples in this figure correspond to the cross-section of the problem's parameter space marked in Fig. \ref{fig1} by the dotted line.}
\label{fig2}
\end{figure}

To find their profiles, boundary-value problem (\ref{3.1}), (\ref{2.19}),
(\ref{2.21}) was solved numerically via the algorithm described in Appendix
\ref{Appendix B}. Examples of the solutions are shown in Fig. \ref{fig2}: one
can see that, as $\rho_{v}\rightarrow\rho_{v.spi}$ (i.e., going from curve 1
to curve 6), the whole solution $\rho(z)$ tends to $\rho_{v}$ -- not just at
the drop's `periphery' (at large $z$), but also in the drop's `core' (finite
$z$). In the opposite limit, $\rho_{v}\rightarrow\rho_{v.sat}$, the drop's
core becomes homogeneous ($\rho\approx\rho_{l.sat}$) and increasingly large;
it is separated from the surrounding vapor by a well-developed interface. One
can say that such solutions describe \emph{macroscopic} drops -- i.e.,
sphere-shaped domains `filled' with liquid.

\begin{figure}
\includegraphics[width=\columnwidth]{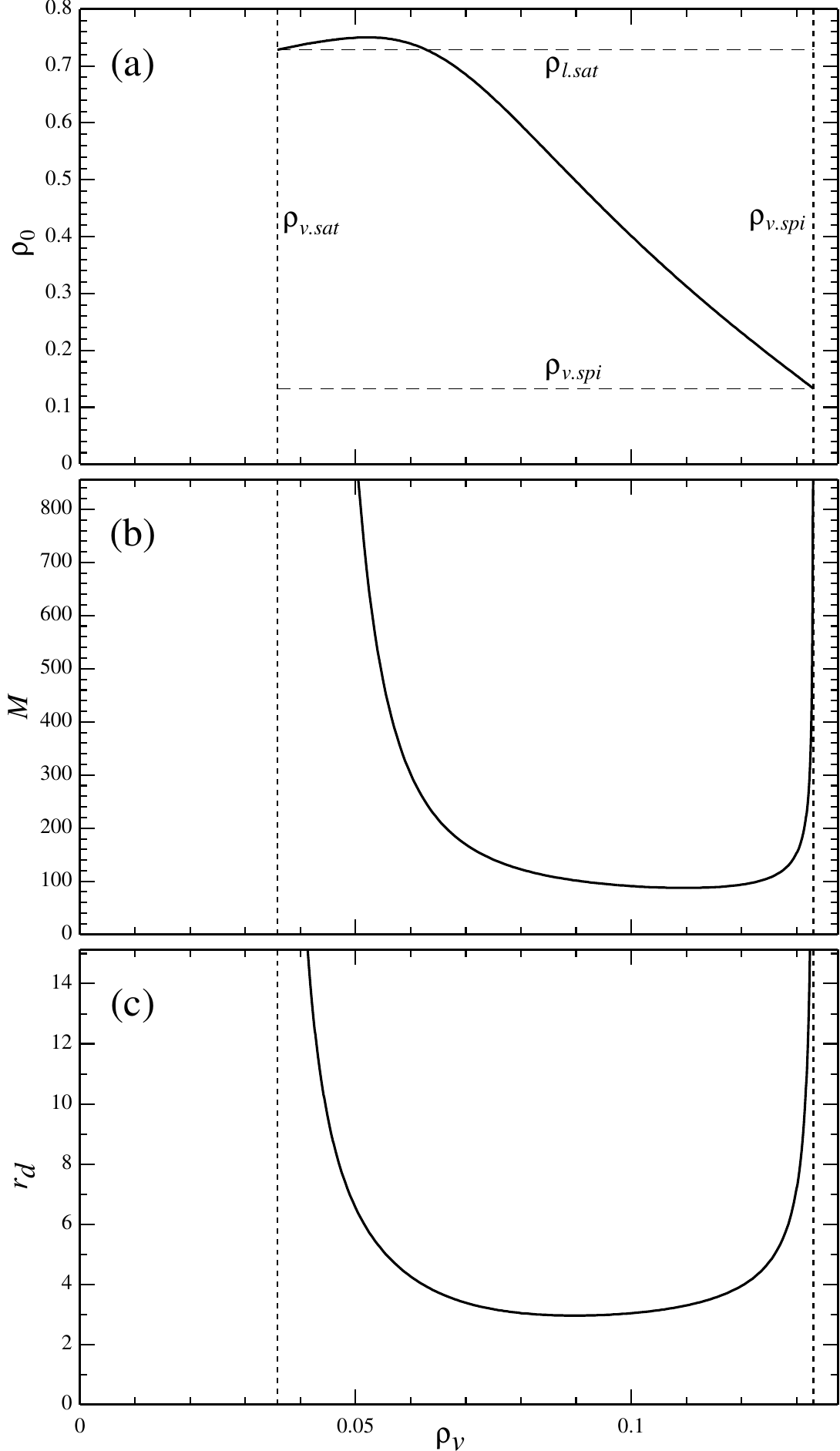}
\caption{The global characteristics of the drop vs vapor density $\rho_{v}$, for $T=0.2$. (a) the density at the drop's center; (b) the drop's excess mass (\ref{3.2}); (c) the drop's radius defined by equation(\ref{3.3}). This figure corresponds to the cross-section of the problem's parameter space marked in Fig. \ref{fig1} by the dotted line.}
\label{fig3}
\end{figure}

There are certain features of the drop profiles that are difficult to see in
Fig. \ref{fig2}, which are however visible in Fig. \ref{fig3}. The latter
illustrates the dependence of the drop's \emph{global} characteristics on the
vapor density $\rho_{v}$.

The following tendencies can be observed:

(1) One would expect the pressure in the drop's center to reach its maximum
when the outside pressure -- i.e., vapor pressure -- does. The same should be
expected for the \emph{density} at the drop's center and the vapor density
(because, in a stable fluid, pressure grows with density). Yet, the opposite
is the case: $\rho_{v}$ reaches its maximum at $\rho_{v}=\rho_{v.spi}$, but
the density at the drop's center, $\rho_{0}=\rho(0)$, is at its minimum there
-- see Fig. \ref{fig3}a. The maximum of $\rho_{0}$ is located near the other
end of the range, $\rho_{v}\approx\rho_{v.sat}$.

(2) Fig. \ref{fig3}b shows the drop's excess mass,%
\begin{equation}
M=4\pi\int_{0}^{\infty}\left(  \rho-\rho_{v}\right)  r^{2}\mathrm{d}%
r,\label{3.3}%
\end{equation}
vs. the vapor density $\rho_{v}$. Since the drop becomes infinitely large in
the limit $\rho_{v}\rightarrow\rho_{v.sat}$, it comes as no surprise that $M$
tends to infinity in this case. It is less clear why $M$ becomes infinite in
the limit $\rho_{v}\rightarrow\rho_{v.spi}$ -- especially since, in this case,
the density at the drop's center becomes indistinguishable from $\rho_{v}$ (as
illustrated in Fig. \ref{fig2}).

(3) The unbounded growth of $M$ as $\rho_{v}\rightarrow\rho_{v.spi}$ is
explained by Fig. \ref{fig3}b which shows the drop's radius $r_{d}$ defined by%
\begin{equation}
\rho(r_{d})=\frac{1}{2}\left(  \rho_{0}+\rho_{v}\right)  . \label{3.4}%
\end{equation}
Evidently, $r_{d}$ grows as the vapor approaches the spinodal point, and this
growth must be strong enough to make $M$ infinite despite the drop's
amplitude, $\rho_{0}-\rho_{v}$, tending to zero.

\section{Evolving drops\label{Sec 4}}

As shown in the previous section, no more than a single static-drop solution
exists for a given vapor density $\rho_{v}$. This implies that all other drops
either evaporate and eventually disappear -- or act as centers of condensation
and grow. It is unclear, however, which of the two scenarios occurs for a
given drop. Another question to be clarified is that of stability of the
steady solutions found in the previous section.

In what follows, these issues are examined for the case of nearly saturated vapor.

\subsection{Large drops in nearly-saturated vapor}

The assumption that the vapor surrounding the drop is near saturated amounts
to%
\begin{equation}
\rho_{v}=\rho_{v.sat}+\delta, \label{4.1}%
\end{equation}
where $\left\vert \delta\right\vert \ll1$ (positive $\delta$ corresponds to
oversaturation and negative, to undersaturation). It will also be assumed that
the drop is large by comparison with the interfacial thickness, i.e.,
$\varepsilon\ll1$.

Under the above assumptions, the profile of the interface separating the drop
and vapor is close to that of a \emph{flat} liquid/vapor interface in an
unbounded space -- i.e., the solution of equations (\ref{2.17})--(\ref{2.18})
can be sought in the form%
\begin{equation}
\rho=\bar{\rho}(z)+\mathcal{O}(\varepsilon,\delta), \label{4.2}%
\end{equation}
where $\bar{\rho}(z)$ is a known function satisfying the flat-interface
boundary-value problem (\ref{2.10})--(\ref{2.13}), the coordinate $z$ is
measured from the current position of the interface,%
\begin{equation}
z=r-\frac{r_{d}}{\varepsilon}, \label{4.3}%
\end{equation}
and $r_{d}(t)$ is the drop's scaled radius (to be determined).

Rewriting equations(\ref{2.17})--(\ref{2.18}) in terms of $z$, one obtains%
\begin{equation}
-\frac{\partial\rho}{\partial z}\frac{\mathrm{d}r_{d}}{\mathrm{d}t}%
+\frac{\partial\left(  v\rho\right)  }{\partial z}=\mathcal{O}(\varepsilon),
\label{4.4}%
\end{equation}%
\begin{multline}
\frac{\partial p}{\partial z}=\varepsilon\frac{\partial}{\partial z}\left[
\left(  \frac{4\mu_{s}}{3}+\mu_{b}\right)  \frac{\partial v}{\partial
z}\right] \\
+\rho\frac{\partial}{\partial z}\left(  \frac{\partial^{2}\rho}{\partial
z^{2}}+\frac{2\varepsilon}{r_{d}}\frac{\partial\rho}{\partial z}\right)
+\mathcal{O}(\varepsilon^{2}), \label{4.5}%
\end{multline}
where the specific form of the higher-order terms (hidden behind the
$\mathcal{O}$ symbols) will not be needed.

Note that the drop's centre $r=0$ corresponds to $z=-r_{d}/\varepsilon$, and
the latter tends to $-\infty$ as $\varepsilon\rightarrow0$. Accordingly,
boundary conditions (\ref{2.19})--(\ref{2.20}) will be replaced with%
\begin{align}
\frac{\partial\rho}{\partial z}  &  \rightarrow0\qquad\text{as}\qquad
z\rightarrow-\infty,\label{4.6}\\
v  &  \rightarrow0\qquad\text{as}\qquad z\rightarrow-\infty. \label{4.7}%
\end{align}
According to ansatz (\ref{4.2})--(\ref{4.3}), the exact solution $\rho$ is
approximated by $\bar{\rho}(z)$ -- and the derivative of $\bar{\rho}(z)$
decays exponentially as $z\rightarrow-\infty$ [see (\ref{2.16})], the error
introduced by moving boundary condition (\ref{4.6}) from a large, but finite,
$z$ to $-\infty$ is exponentially small. The same conclusion can be reached
for condition (\ref{4.7}) using the solution for $v$ calculated below.

Rewriting boundary conditions (\ref{2.21})--(\ref{2.22}) in terms of $z$ is a
straightforward matter,%
\begin{align}
\rho &  \rightarrow\rho_{v}\qquad\text{as}\qquad z\rightarrow\infty
,\label{4.8}\\
\frac{\partial v}{\partial z}  &  \rightarrow0\qquad~\,\text{as}\qquad
z\rightarrow\infty. \label{4.9}%
\end{align}
The main characteristic of the drop is its radius $r_{d}(t)$. To find it, one
should expand the solution of equations (\ref{4.4})--(\ref{4.9}) in
$\varepsilon$ and $\delta$, ensure that substitution (\ref{4.2}) satisfies
them to the zeroth order, and examine the first-order equations. As usual,
these have a solution only subject to a certain orthogonality condition which
yields a differential equation for $r_{d}(t)$.

This equation, however, can be derived in a simpler way: by manipulating the
exact equations into a single equality with the zeroth-order terms canceled
out. The resulting \emph{first}-order equality yields the equation for
$r_{d}(t)$.

To see this plan through, multiply equation(\ref{4.5}) by $\left(  \rho
_{0}-\rho\right)  $ where $\rho_{0}$ is the limiting density as $z\rightarrow
-\infty$ (physically, it corresponds to the density in the drop's center), and
integrate from $z=-\infty$ to $z=+\infty$. Using identity (\ref{2.7}) to
replace $\left(  \rho_{0}/\rho\right)  \left(  \partial p/\partial r\right)
\rightarrow\rho_{0}\partial G/\partial r$, integrating the first integral on
the right-hand side by parts, and using boundary conditions (\ref{4.6}%
)--(\ref{4.9}), one obtains%
\begin{multline}
\rho_{0}\left[  G(\rho_{v})-G(\rho_{0})\right]  -p(\rho_{v})+p(\rho_{0})\\
=\varepsilon\int_{0}^{\infty}\frac{\rho_{0}}{\rho^{2}}\frac{\partial\rho
}{\partial r}\left(  \frac{4\mu_{s}}{3}+\mu_{b}\right)  \frac{\partial
v}{\partial r}\mathrm{d}r\\
+\varepsilon\int_{0}^{\infty}\frac{2}{r}\left(  \frac{\partial\rho}{\partial
r}\right)  ^{2}\mathrm{d}r+\mathcal{O}(\varepsilon^{2}). \label{4.10}%
\end{multline}
This is the desired equality without the leading order.

To ascertain that all zeroth-order terms have indeed cancelled out, observe
that the nearly-flat-interface ansatz (\ref{4.2}) implies that the density at
the drop's center is close to that of saturated liquid,%
\[
\rho_{0}=\rho_{l.sat}+\mathcal{O}(\delta).
\]
Using this equality, as well as (\ref{4.1}) and (\ref{2.7}), one can rearrange
the left-hand side of equation (\ref{4.10}) in the form%
\[
lhs~of~\left(  \ref{4.10}\right)  =\frac{\rho_{l.sat}-\rho_{v.sat}}%
{\rho_{v.sat}}\left(  \frac{\partial p}{\partial\rho}\right)  _{\rho
=\rho_{v.sat}}\delta+\mathcal{O}(\delta^{2}).
\]
Evidently, all zeroth-order terms have indeed cancelled out from this
expression, and so equation (\ref{4.10}) becomes%
\begin{multline}
\delta\frac{\rho_{l.sat}-\rho_{v.sat}}{\rho_{v.sat}}\left(  \frac{\partial
p}{\partial\rho}\right)  _{\rho=\rho_{v.sat}}\\
=\varepsilon\int_{0}^{\infty}\frac{\rho_{0}}{\rho^{2}}\left(  \frac{4}{3}%
\mu_{s}+\mu_{b}\right)  \frac{\partial\rho}{\partial r}\frac{\partial
v}{\partial r}\mathrm{d}r\\
+\varepsilon\int_{0}^{\infty}\frac{2}{r}\left(  \frac{\partial\rho}{\partial
r}\right)  ^{2}\mathrm{d}r+\mathcal{O}(\varepsilon^{2},\delta^{2}%
).\label{4.11}%
\end{multline}
To close equation (\ref{4.11}), it remains to express $v$ in terms of
$\bar{\rho}$. To do so, substitute ansatz (\ref{4.2}) into equation
(\ref{4.4}) and, taking into account (\ref{4.7}) and (\ref{4.9}), obtain%
\[
v=\left(  1-\frac{\rho_{l.sat}}{\bar{\rho}}\right)  \frac{\mathrm{d}r_{d}%
}{\mathrm{d}t}+\mathcal{O}(\varepsilon).
\]
Substituting this expression into equation (\ref{4.11}) and omitting small
terms, one obtains%
\begin{equation}
\frac{\mathrm{d}r_{d}}{\mathrm{d}t}=D-\frac{2\sigma}{Ar_{d}},\label{4.12}%
\end{equation}
where%
\begin{equation}
\sigma=\int_{-\infty}^{\infty}\left(  \frac{\mathrm{d}\bar{\rho}}{\mathrm{d}%
z}\right)  ^{2}\mathrm{d}z,\label{4.13}%
\end{equation}%
\begin{equation}
A=\rho_{l.sat}^{2}\int_{-\infty}^{\infty}\frac{\frac{4}{3}\mu_{s}(\bar{\rho
})+\mu_{b}(\bar{\rho})}{\bar{\rho}^{4}}\left(  \frac{\mathrm{d}\bar{\rho}%
}{\mathrm{d}z}\right)  ^{2}\mathrm{d}z,\label{4.14}%
\end{equation}%
\[
D=\left(  \frac{1}{\rho_{v.sat}}-\frac{1}{\rho_{l.sat}}\right)  \left(
\frac{\partial p}{\partial\rho}\right)  _{\rho=\rho_{v}}\frac{\delta
}{\varepsilon}.
\]
Physically, $\sigma$ is the nondimensional surface tension. The coefficient
$A$, in turn, was introduced in Refs. \cite{Benilov20d,Benilov22b} as one of
the characteristics of the flow due to evaporation or condensation near an
almost flat interface. This is what $A$ characterizes in the present problem
for the particular case of a spherical interface. The coefficient $D$, in
turn, characterizes the degree of over/under-saturation of the vapor
surrounding the drop.

\subsection{Solutions of equation (\ref{4.12})}

\begin{figure}
\includegraphics[width=\columnwidth]{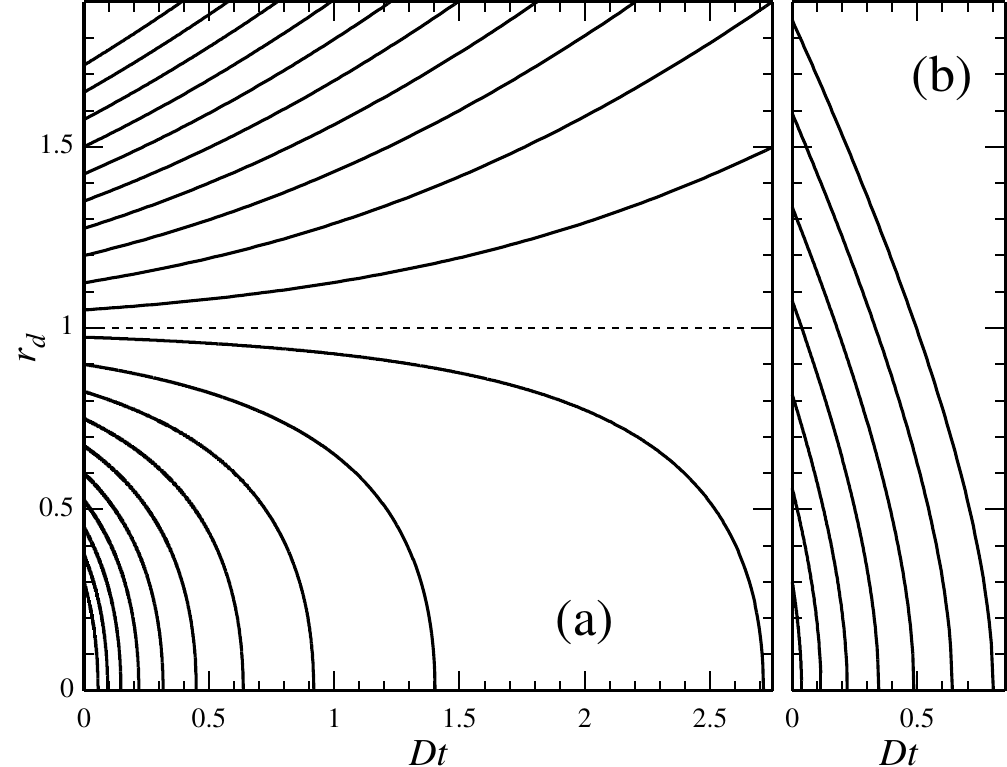}
\caption{Solution (\ref{4.16}) with various initial conditions for (a) $2\sigma/AD=1$, (b) $2\sigma/AD=-1$. The horizontal dotted line in panel (a) shows the separatrix $r_{d}=2\sigma/AD$.}
\label{fig4}
\end{figure}

The general solution of equation (\ref{4.12}) can be readily found in an
implicit form,%
\begin{equation}
\dfrac{2\sigma}{AD}\ln\frac{\dfrac{2\sigma}{AD}-r_{d}(t)}{\dfrac{2\sigma}%
{AD}-r_{d}(0)}+r_{d}(t)-r_{d}(0)=Dt, \label{4.15}%
\end{equation}
where $r_{d}(0)$ is the initial condition. Typical solutions for $D>0$
(oversaturated vapor) are shown in Fig. \ref{fig4}a. One can see [and confirm
by examining equation (\ref{4.15})] that there are two patterns of evolution:
smaller drops evaporate, whereas larger drops absorb surrounding oversaturated
vapor and grow\footnote{It can be shown that the drops that grow do so
linearly, i.e., $r_{d}=Dt+\mathcal{O}(\ln t)$ as $t\rightarrow\infty.$}. The
(steady) solution corresponding to the separatrix,%
\[
r_{d}=\frac{2\sigma}{AD}\qquad\text{for all }t,
\]
corresponds to the static drop-solution computed in Sec. \ref{Sec 3} (or, more
precisely, its weakly-oversaturated limit).

For $D<0$ (undersaturated vapor), all drops evaporate -- see examples of
solutions depicted in Fig. \ref{fig4}b. Finally, in the limit $D\rightarrow0$
(saturated vapor), (\ref{4.15}) yields%
\[
\frac{A}{4\sigma}\left[  r_{d}^{2}(0)-r_{d}^{2}(t)\right]  =t.
\]
This dependence of the radius of an evaporating drop on time is often referred
to as the \textquotedblleft$d^{2}$-law\textquotedblright\ (e.g.,
\cite{ErbilMchaleNewton02,SaxtonWhiteleyVellaOliver16,DallabarbaWangPicano21}%
). Its single most important characteristic is the time of full evaporation of
a drop of radius $r_{d}(0)$,%
\begin{equation}
t_{e}=\frac{Ar_{d}^{2}(0)}{4\sigma}. \label{4.16}%
\end{equation}
To make $t_{e}$ easier to calculate, expressions (\ref{4.13})--(\ref{4.14})
for the coefficients $\sigma$ and $A$ will be rewritten as closed-form
integrals [in their current form, they involve the solution $\bar{\rho}(z)$ of
a `third-party' boundary-value problem (\ref{2.15}), (\ref{2.11})--(\ref{2.13})].

To do so, multiply equation (\ref{2.15}) by $\mathrm{d}\bar{\rho}/\mathrm{d}%
z$, integrate it, use identity (\ref{2.7}), and fix the constant of
integration via condition (\ref{2.11}), which yields%
\[
\frac{\mathrm{d}\bar{\rho}}{\mathrm{d}z}=-\sqrt{2\left[  \bar{\rho}G(\bar
{\rho})-p(\bar{\rho})-\bar{\rho}G(\rho_{v.sat})+p(\rho_{v})\right]  },
\]
where it has been taken into account that $\bar{\rho}(z)$ is a monotonically
decreasing function.

Using the above equation and boundary conditions (\ref{2.11})--(\ref{2.12})
and omitting the overbar from $\bar{\rho}$, one can rearrange expressions
(\ref{4.13})--(\ref{4.14}) in the form\begin{widetext}%
\begin{equation}
\sigma=\int_{\rho_{v.sat}}^{\rho_{l.sat}}\sqrt{2\left[  \rho G(\rho
)-p(\rho)-\rho G(\rho_{v.sat})+p(\rho_{v.sat})\right]  }\mathrm{d}%
\rho,\label{4.17}%
\end{equation}%
\begin{equation}
A=\rho_{l.sat}^{2}\int_{\rho_{v.sat}}^{\rho_{l.sat}}\frac{\mu(\rho)}{\rho^{4}%
}\sqrt{2\left[  \rho G(\rho)-p(\rho)-\rho G(\rho_{v.sat})+p(\rho
_{v.sat})\right]  }\mathrm{d}\rho,\label{4.18}%
\end{equation}
\end{widetext}where%
\begin{equation}
\mu(\rho)=\frac{4\mu_{s}(\rho)}{3}+\mu_{b}(\rho) \label{4.19}%
\end{equation}
is the effective viscosity.

\subsection{Stability of steady solutions}

The mere fact that the steady-drop solutions computed in Sec. \ref{Sec 3} act
as separatrices between two sets of diverging trajectories (as illustrated in
Fig. \ref{fig4}a) makes these solutions unstable.

Indeed, consider an initial condition involving a steady drop plus a small
perturbation. If the latter has positive mass, the drop immediately starts
growing further, whereas a perturbation with a negative mass triggers off
evaporation. In either case, the solution evolves \emph{away} from the
unperturbed steady state, which amounts to instability.

To interpret it physically, recall that the steady solution exists only for
drops surrounded by \emph{over}saturated vapor -- which is stable with respect
to small perturbations, but may be unstable with respect to finite ones. One
can thus assume that the steady-drop solution represents the `minimal'
perturbation capable of triggering off instability of the surrounding vapor.

It should be emphasized that all of the results and conclusions in this
section have been obtained for the asymptotic limit $0<\delta\ll1$ (slightly
oversaturated vapor), but the clear physical interpretation suggests that
\emph{all} steady-drop solutions are unstable, not only those with small
$\delta$.

\subsection{The Hertz--Knudsen equation}

The evaporative flux $\mathcal{F}$ from or to a drop floating in vapor is,
essentially, the left-hand side of equation (\ref{4.12}) multiplied by $4\pi
r_{d}^{2}\,\rho_{l.sat}$. Thus, multiplying the right-hand side by the same
factor, one can rearrange it in the form%
\[
\mathcal{F}=\frac{4\pi r_{d}^{2}\,\rho_{l.sat}}{A}\left(  DA-\frac{2\sigma
}{r_{d}}\right)  .
\]
This expression has essentially the same structure as the semi-empiric
Hertz--Knudsen equation \cite{Hertz82,Knudsen15}: the first term in brackets
is the pressure difference due to oversaturation and the second term, the
pressure difference due to capillarity.

\section{Evaporation of water drops in saturated vapor\label{Sec 5}}

In this section, the model described above will be illustrated by calculating
the evaporation times of realistic water drops floating in water vapor. For
simplicity, the latter will be assumed to be saturated.

Before carrying out the calculations, one needs to estimate the model's
parameters by linking them to experimental data.

\subsection{The effective viscosity $\mu(\rho)$}

To calculate the coefficient $A$, one has to specify $\mu(\rho)$. In this
work, the simplest approximation is assumed:%
\begin{equation}
\mu=\mu_{0}\rho^{n}, \label{5.1}%
\end{equation}
where $\mu_{0}$ and $n$ depend on $T$.

To estimate $\mu_{0}$ and $n$ for water at, say, $25^{\circ}\mathrm{C}$, note
that the densities and shear viscosities of the water vapor and liquid are
\cite{LindstromMallard97}%
\[
\rho_{v.sat}=0.023075\,\mathrm{kg}/\mathrm{m}^{3},\hspace{0.5cm}\rho
_{l.sat}=997.00\,\mathrm{kg}/\mathrm{m}^{3},
\]%
\[
\mu_{s}(\rho_{v.sat})=9.8669\times10^{-6}\mathrm{Pa\,s},
\]%
\[
\mu_{s}(\rho_{l.sat})=0.00089011\,\mathrm{Pa\,s}.
\]
Measurements of the bulk viscosity, in turn, are scarce; the author of the
present paper was able to find those only for the liquid water
\cite{HolmesParkerPovey11},%
\[
\mu_{b}(\rho_{l.sat})=2.47\times10^{-3}\mathrm{Pa\,s}.
\]
For lack of a better option, the bulk viscosity of vapor was estimated via a
`proportionality hypothesis',%
\[
\frac{\mu_{b}(\rho_{v.sat})}{\mu_{b}(\rho_{l.sat})}=\frac{\mu_{s}(\rho
_{v.sat})}{\mu_{s}(\rho_{l.sat})},
\]
suggesting that%
\[
\mu_{b}(\rho_{v.sat})=2.738\times10^{-5}\mathrm{Pa\,s}.
\]
Recalling definition (\ref{4.19}) of $\mu$ and fitting dependence (\ref{5.1})
to the above values, one obtains%
\begin{equation}
\mu_{0}\approx21.635,\qquad n\approx0.42180. \label{5.2}%
\end{equation}

\subsection{The coefficients $\sigma(T)$ and $A(T)$}

\begin{figure}
\includegraphics[width=\columnwidth]{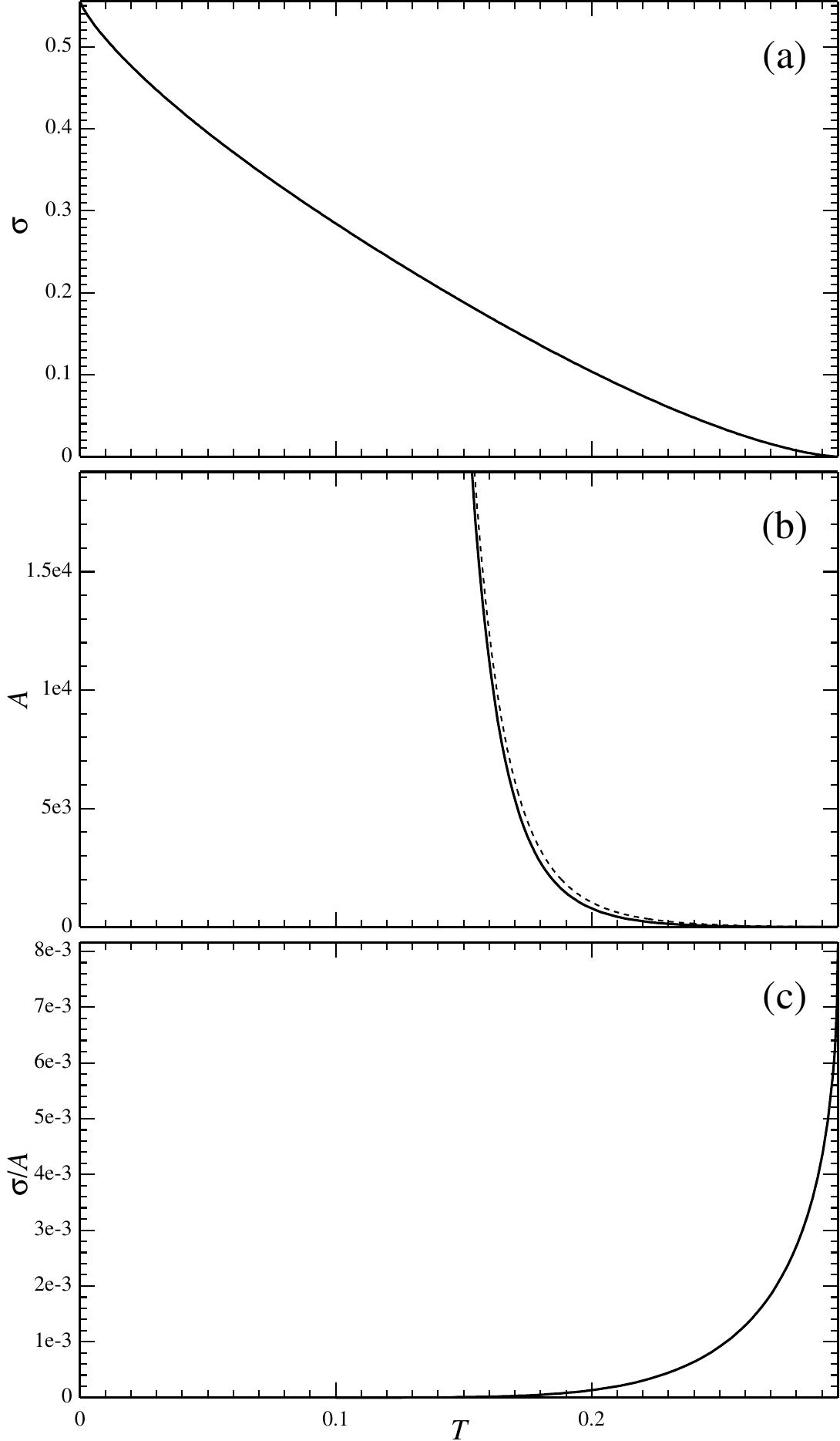}
\caption{The dependence of the coefficients of equation (\ref{4.12}) on the nondimensional temperature: (a) $\sigma(T)$, (b) $A(T)$, (c) $\sigma(T)/A(T)$. The dotted curve in panel (b) corresponds to asymptotic expression (\ref{5.3}).}
\label{fig5}
\end{figure}

Fig. \ref{fig5} shows plots of $\sigma$, $A$, and $\sigma/A$ vs. $T$, computed
for the van der Waals fluid and $\mu(\rho)$ given by (\ref{5.1})--(\ref{5.2}).
The following two features can be observed.

(1) In the limit of near-critical temperature $T\rightarrow T_{cr}$ both
$\sigma$ and $A$ vanish. This occurs because%
\[
\rho_{v.sat}\rightarrow\rho_{cr}\leftarrow\rho_{l.sat}\qquad\text{as}\qquad
T\rightarrow T_{cr},
\]
and so the integration intervals in (\ref{4.17})--(\ref{4.18}) shrink to a point.

Note that the ratio $\sigma/A$ remains finite in this limit, as expressions
(\ref{4.17})--(\ref{4.18}) imply%
\[
\frac{\sigma}{A}\rightarrow\frac{\mu(\rho_{cr})}{\rho_{cr}^{2}}\qquad
\text{as}\qquad T\rightarrow T_{cr}.
\]
In fact, $\sigma/A$ has a maximum at the critical point (as Fig. \ref{fig5}c
suggests), so that the Kelvin-effect-induced evaporation is at its strongest.

(2) If $\rho_{v.sat}$ is small, the factor $\mu/\rho^{4}$ in the integrand of
(\ref{4.18}) becomes large near the lower limit. As a result, one can find $A$ approximately.

Since the main contribution to integral (\ref{4.18}) comes from the region
with small $\rho$, the fluid's equations of state can be approximated by that
of ideal gas,%
\[
p=T\rho+\mathcal{O}(\rho^{2}),\qquad G=T\ln\rho+\operatorname{const}%
+\mathcal{O}(\rho),
\]
where the value of $\operatorname{const}$ is unimportant and the other
coefficients are implied to have been eliminated during nondimensionalization.
Substituting approximation (\ref{5.1}) for $\mu$ and the ideal-gas
approximations of $p$ and $G$ into expression (\ref{4.18}) for $A$, changing
$\rho\rightarrow\xi=\rho/\rho_{v.sat}$, and keeping the leading-order terms
only, one obtains%
\[
A\approx\frac{\mu_{0}T^{1/2}}{\rho_{v.sat}^{5/2-n}}\int_{1}^{\infty}%
\frac{2^{1/2}}{\xi^{4-n}}\sqrt{\xi\ln\xi-\xi+1}\mathrm{d}\xi.
\]
The integral in this expression can be evaluated numerically: assuming that
$n$ is given by (\ref{5.2}), one obtains%
\begin{equation}
A\approx\frac{4.3597\,T^{1/2}}{\rho_{v.sat}^{2.0782}}. \label{5.3}%
\end{equation}
This result is compared to the exact one in Fig. \ref{fig5}b.

As for the small-$\rho_{v.sat}$ limit of $\sigma$, it cannot be calculated
without specifying the equation of state. It can be safely assumed, however,
that $\sigma$ remains finite as $\rho_{v.sat}\rightarrow0$ (for the van der
Waals fluid, for example, $\sigma\rightarrow2^{-5/2}\pi$). Thus, the
right-hand side of equation (\ref{4.12}) tends to zero in this limit, so that
the Kelvin-effect-induced evaporation is weak.

Since $\rho_{v}$ is small even for moderately small $T$ (due to the
exponential nature of their relationship \cite{Benilov20d}), the above
conclusion applies to both small and moderate temperatures (see Fig.
\ref{fig5}c).

\subsection{An estimate of drop evaporation time}

\begin{figure}
\includegraphics[width=\columnwidth]{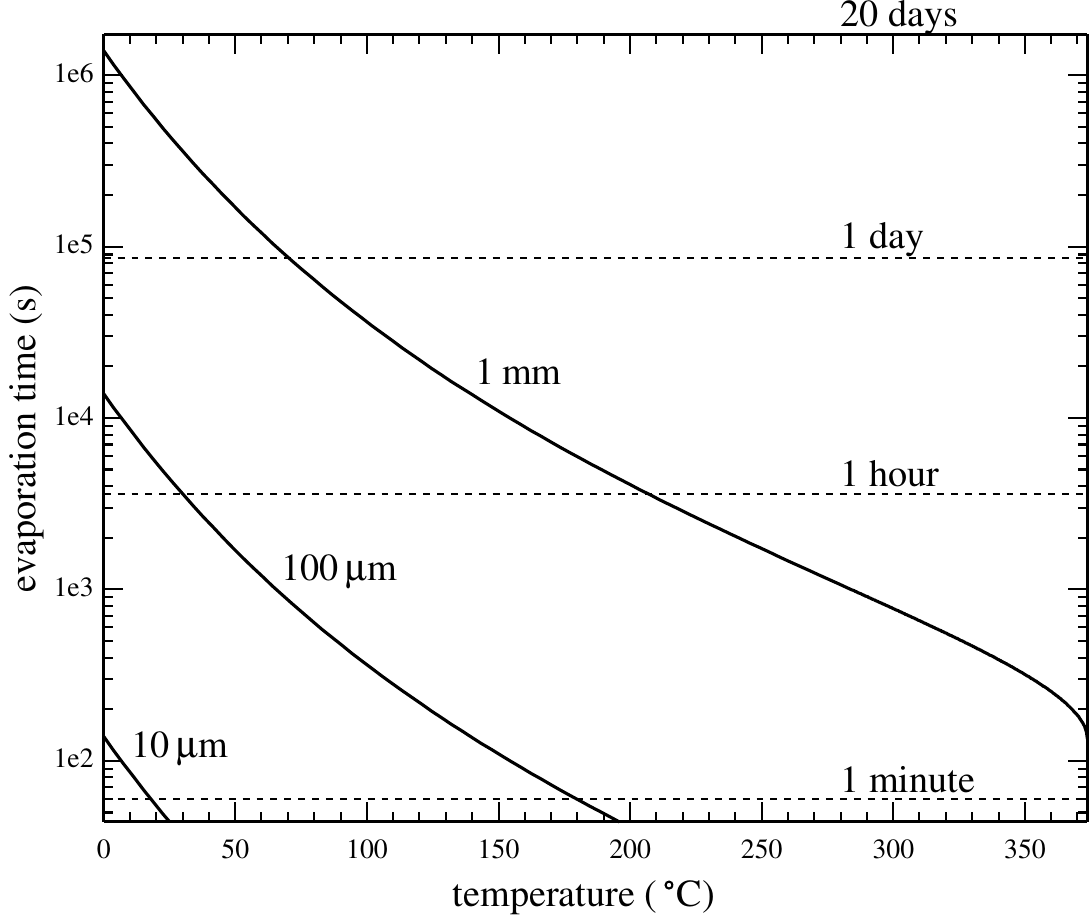}
\caption{The drop evaporation time vs. temperature. The value of the drop's initial radius are indicated near the corresponding curve.}
\label{fig6}
\end{figure}

To find out how significant the Kelvin effect is for evaporation of drops, the
\emph{dimensional} evaporation time has been calculated for several drop
sizes. For simplicity, the van der Waals equation of state was used, with the
parameters of water estimated in Ref. \cite{HaynesLideBruno17},%
\[
a=1704.8\,\mathrm{m}^{5}/\mathrm{kg}\,\mathrm{s}^{2},\qquad
b=0.0016918\,\mathrm{m}^{3}/\mathrm{kg}.
\]
Thus, the density and pressure scales used in the nondimensionalization are%
\[
\varrho=\frac{1}{b}=591.09\,\mathrm{kg}/\mathrm{m}^{3},\qquad P=\frac{a}%
{b^{2}}=5956.3\,\mathrm{bar}.
\]
For the Korteweg parameter, the following estimate \cite{Benilov20b} was
assumed:%
\[
K=0.890\times10^{-16}\,\mathrm{m}^{7}/\mathrm{kg}\,\mathrm{s}^{2}.
\]
The dimensional equivalent of expression (\ref{4.16}) for the drop evaporation
time is plotted in Fig. \ref{fig6} vs. the dimensional temperature, varying
between $0.01^{\circ}\mathrm{C}$ (the triple point) and $373.95^{\circ
}\mathrm{C}$ (the critical point). Evidently, millimeter-sized drops at under
$70^{\circ}\mathrm{C}$ survive for days, whereas smaller drops evaporate much
quicker even at low temperatures. It should be added that micron-size drops
(not shown in Fig. \ref{fig6}) evaporate in under $1.6\,\mathrm{s}$ for the
whole temperature range.

When comparing the results presented in Fig. \ref{fig6} with the predictions
of one's intuition, one should keep in mind that the former is for
\emph{vapor}, whereas the latter is trained by one's experience with
\emph{air}. The parameters of these two gases differ dramatically: the density
of water vapor at, say, $25^{\circ}\mathrm{C}$ is $0.023075$ $\mathrm{kg}%
/\mathrm{m}^{3}$ \cite{LindstromMallard97} -- whereas that of air (at the same
temperature and pressure of $1\,\mathrm{atm}$) is $1.1699$ $\mathrm{kg}%
/\mathrm{m}^{3}$ \cite{CzerniaSzyk21}, with the viscosity difference being
probably just as large.

Thus, to accurately predict evaporation times, one needs to redo the present
calculation for a multicomponent fluid, and also an equation of state, more
accurate than the van der Waals one. The present work should rather be viewed
as a proof of concept.

\section{Concluding remarks}

Thus, a model of evaporation and condensation of liquid drops surrounded by
vapor of the same fluid has been presented. It was shown that, in saturated
vapor, all drops evaporate (due to the Kelvin effect), but a steady solution
describing a static drop exists for each vapor density between the saturated
one $\rho_{v.sat}$ and the spinodal one $\rho_{v.spi}$. These static solutions
act as separatrices: smaller drops evaporate, larger drops act as a center of
condensation and grow. Strictly speaking, this conclusion was obtained for
\emph{weakly}-oversaturated vapor, but is likely to hold in the whole range
where static drops exist, i.e., between $\rho_{v.sat}$ and $\rho_{v.spi}$.

Both evaporation and condensation occur because a drop of an arbitrary radius
cannot be in both mechanical and thermodynamic equilibria with the surrounding
vapor. As a result, a flow develops -- either from the drop toward the vapor
or vice versa. This mechanism differs from the one based on diffusion, assumed
previously for drops surrounded by air. Since the `new' mechanism is also
present in the `old' setting -- yet not included in the `old' models -- the
next step should be an extension of the present results to multicomponent
fluids. With this done, one should be able to compare the theoretical and
experimental results.

Another shortcoming of the present model -- albeit an easily curable one -- is
approximation (\ref{5.1}) of the fluid viscosity. Even though it should work
reasonably well in the general case, it may not do so when $\rho_{v.sat}$ is
small and the vapor can be viewed as a diluted gas. In this limit, both
kinetic theory (e.g., \cite{FerzigerKaper72}) and measurements (e.g.,
\cite{LindstromMallard97}) suggest that the dynamic viscosities $\mu_{s}$ and
$\mu_{b}$ do not depend on the density and can be treated as constants. As a
result, the small-$\rho_{v.sat}$ expression (\ref{5.3}) for the coefficient
$A$ no longer applies. Its correct version can be calculated in the same
manner, which yields%
\[
A\approx\frac{0.14219\,T^{1/2}\left[  \frac{4}{3}\mu_{s}(\rho_{v.sat})+\mu
_{b}(\rho_{v.sat})\right]  }{\rho_{_{v.sat}}^{5/2}}.
\]
This correction also implies that the low-$T$ (small-$\rho_{v.sat}$) end of
Fig. \ref{fig6} needs to be recalculated with the new approximation of the viscosity.

Still, even the present version of the model can be used to experimentally
prove the existence of the Kelvin-effect-induced evaporation and condensation
(to the best of the author's knowledge, this has yet to be done).

One way of doing so consists in filling an air-tight container with a
single-component vapor, at a near-critical temperature (in which case the
Kelvin effect is at its strongest). Two narrow cylindrical vessels, partially
filled with liquid, should be placed in the container -- one made of a
hydrophobic material and the other, of a hydrophilic one. After a certain
time, however, all of the liquid should end up in the hydrophilic vessel --
simply because the liquid/vapor interface in it is concave (hence, Kelvin
effect gives rise to condensation), whereas the convex interface in the
hydrophobic vessel is conducive to evaporation.

\section{Author declarations}

The author has no conflicts to disclose.

\section{Data availability}

The data that support the findings of this study are available within the article.

\appendix

\section{When does boundary-value problem (\ref{3.1}), (\ref{2.19}),
(\ref{2.21}) have solutions?\label{Appendix A}}

Physically, the function $G(\rho)$ representing a fluid's chemical potential
should have the following properties (see a schematic in Fig. \ref{fig7}a):

\begin{figure}
\includegraphics[width=\columnwidth]{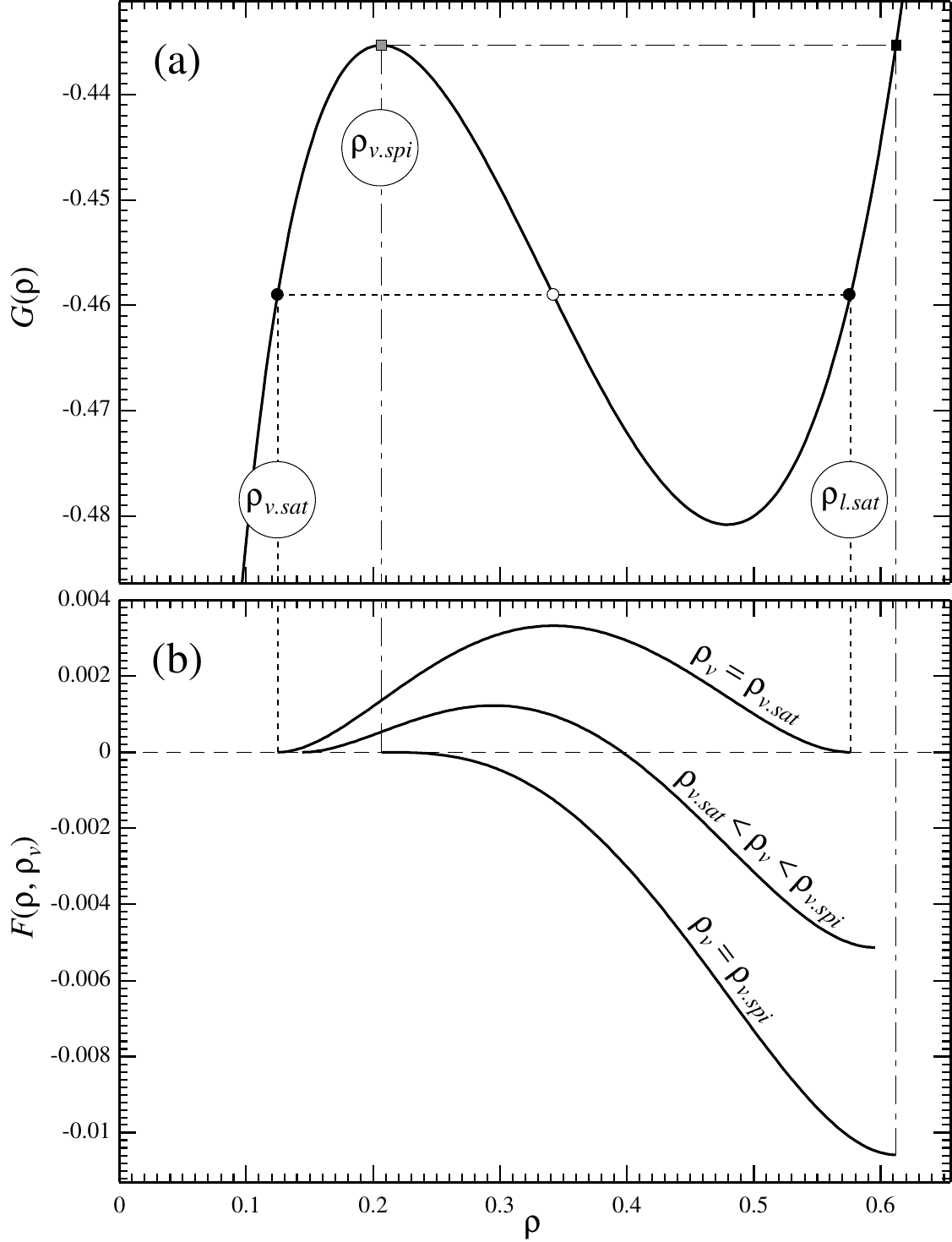}
\caption{Thermodynamic functions for the van der Waals fluid (\ref{2.8})--(\ref{2.9}) with $T=0.26$: (a) chemical potential $G(\rho)$, (b) the function $F(\rho)$ defined by (\ref{A.9}).}
\label{fig7}
\end{figure}

\begin{itemize}
\item In the low-density limit, $G(\rho)$ should tend to the chemical
potential of an ideal gas,%
\[
G\sim T\ln\rho\qquad\text{as}\qquad\rho\rightarrow0.
\]

\item $G(\rho)$ should have a local maximum at $\rho_{v.spi}$ and a local
minimum at $\rho_{l.spi}>\rho_{v.spi}$ (the latter is not labelled in Fig.
\ref{fig7}a).

\item $G(\rho)$ should tend to infinity either as $\rho\rightarrow\infty$ or
at a finite point $\rho\rightarrow\rho_{\infty}$. (The van der Waals fluid
(\ref{2.9}), for example, is of the latter kind, with $\rho_{\infty}=1$).
\end{itemize}

Now, consider a given value of $\rho_{v}$ [which appears in boundary condition
(\ref{2.21})] and define $\rho_{i}$ and $\rho_{l}$ such that%
\begin{equation}
G(\rho_{v})=G(\rho_{i})=G(\rho_{l}), \label{A.1}%
\end{equation}
and%
\[
\rho_{v}<\rho_{i}<\rho_{l}.
\]
These three values can be visualized in Fig. \ref{fig7}a as the points of
intersection between the graph of $G(\rho)$ and a horizontal line cutting it
at the level of $G(\rho_{v})$. Evidently, if $\rho_{v}$ happens to coincide
with $\rho_{v.sat}$, then $\rho_{l}$ coincides with $\rho_{l.sat}$ and
$\rho_{i}$, with a value marked by a small empty circle.

The answer to the title question of this appendix will be obtained using the
following three lemmas.\medskip

\emph{Lemma 1.} The boundary-value problem (\ref{3.1}), (\ref{2.19}),
(\ref{2.21}) may only admit solutions such that%
\[
\rho_{v}\leq\rho(r)\leq\rho_{l}\qquad\text{for all }r\text{.}%
\]
\emph{Proof.} Linearizing equation (\ref{3.1}) at large $r$ and recalling
boundary condition (\ref{2.21}), one can readily show that%
\begin{equation}
\rho\sim\rho_{v}+\frac{C_{v}}{r}\operatorname{e}^{-\lambda r}\qquad
\text{as}\qquad r\rightarrow\infty, \label{A.2}%
\end{equation}
where $C_{v}$ is an undetermined constant and%
\begin{equation}
\lambda=\sqrt{\left(  \frac{\partial G}{\partial\rho}\right)  _{\rho=\rho_{v}%
}}. \label{A.3}%
\end{equation}
Note that $C_{v}=0$ corresponds to the trivial solution, $\rho=\rho_{v}$ for
all $r$, and so this case will not be considered.

Next, rewrite (\ref{3.1}), (\ref{2.19}), and (\ref{2.21}) in terms of
$\xi=r^{-1}$,%
\begin{equation}
\frac{\mathrm{d}^{2}\rho}{\mathrm{d}\xi^{2}}=\frac{G(\rho)-G(\rho_{v})}%
{\xi^{4}}, \label{A.4}%
\end{equation}%
\begin{equation}
\rho\rightarrow\rho_{v}\qquad\text{as}\qquad\xi\rightarrow0, \label{A.5}%
\end{equation}%
\begin{equation}
\frac{\mathrm{d}\rho}{\mathrm{d}\xi}\rightarrow0\qquad\text{as}\qquad
\xi\rightarrow\infty. \label{A.6}%
\end{equation}
Unlike the original equation (\ref{3.1}), (\ref{A.4}) allows one to readily
determine the sign of the solution's curvature. Recalling definition
(\ref{A.1}) of $\rho_{l}$ and $\rho_{i}$, and looking at the right-hand side
of equation (\ref{A.4}), one can deduce that%
\begin{align*}
\frac{\mathrm{d}^{2}\rho}{\mathrm{d}\xi^{2}}  &  <0\qquad\text{if
\hspace{1.39cm}}\rho<\rho_{v},\\
\frac{\mathrm{d}^{2}\rho}{\mathrm{d}\xi^{2}}  &  >0\qquad\text{if}\qquad
\rho_{v}<\rho<\rho_{i},\\
\frac{\mathrm{d}^{2}\rho}{\mathrm{d}\xi^{2}}  &  <0\qquad\text{if}\qquad
\rho_{i}<\rho<\rho_{l},\\
\frac{\mathrm{d}^{2}\rho}{\mathrm{d}\xi^{2}}  &  >0\qquad\text{if\hspace
{1.46cm}}\rho>\rho_{l}.
\end{align*}
Furthermore, it can be deduced from (\ref{A.2}) that the small-$\xi$
asymptotics of $\rho(\xi)$ is%
\[
\rho\sim\rho_{v}+C_{v}\xi\operatorname{e}^{-\lambda/\xi}\qquad\text{as}%
\qquad\xi\rightarrow0.
\]
For $\xi>0$, three possible behaviors of $\rho(\xi)$ can be identified (as
illustrated in Fig. \ref{fig8}):

\begin{figure}
\includegraphics[width=\columnwidth]{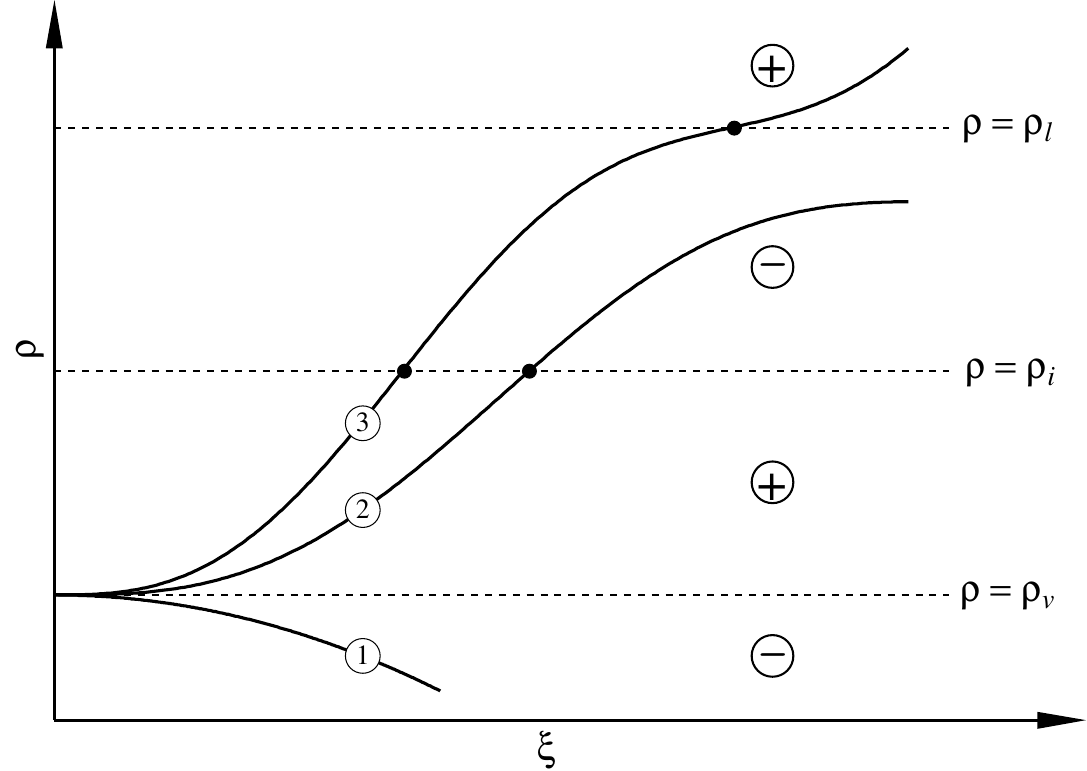}
\caption{Possible solutions of boundary-value problem (\ref{A.4})--(\ref{A.6}). The sign of $\mathrm{d}^{2}\rho/\mathrm{d}\xi^{2}$ [deduced from the right-hand side of equation (\ref{A.4}) and definition (\ref{A.1}) of $\rho_{l}$ and $\rho_{i}$] is indicated within the corresponding band of $\rho$. The inflection points are marked with black dots.}
\label{fig8}
\end{figure}

(i) The solutions emerges from $\xi=0$ with $C_{v}<0$ and stays below
$\rho_{v}$ for a certain range of $\xi$.

(ii) The solution emerges from $\xi=0$ with $C_{v}>0$ and stays inside the
strip $\rho_{v}\leq\rho\leq\rho_{l}$ for all $\xi$.

(iii) The solution emerges from $\xi=0$ with $C_{v}>0$, crosses the strip, and
stays above $\rho_{l}$ for a certain range of $\xi$.

Evidently, a solution with behavior (i) can satisfy boundary condition
(\ref{A.6}) only if it changes its curvature -- which is, however, impossible
as there are no inflection points with $\rho<\rho_{v}$. Similarly, behavior
(iii) is impossible due to the absence of inflection points with $\rho
>\rho_{l}$.

Thus, the only possible behavior is (ii), QED.\medskip

Next, define a function $F(\rho,\rho_{v})$ such that%
\begin{equation}
\frac{\partial F(\rho,\rho_{v})}{\partial\rho}=G(\rho)-G(\rho_{v}),
\label{A.7}%
\end{equation}%
\begin{equation}
F(\rho,\rho_{v})=0\qquad\text{if}\qquad\rho=\rho_{v}. \label{A.8}%
\end{equation}
Using identity (\ref{2.7}), one can deduce that%
\begin{equation}
F(\rho,\rho_{v})=\rho\left[  G(\rho)-G(\rho_{v})\right]  -p(\rho)+p(\rho_{v}).
\label{A.9}%
\end{equation}
The following lemma can be readily verified using equalities (\ref{A.7}%
)--(\ref{A.8}) and the properties of $G(\rho)$ listed in the beginning of this
appendix.\medskip

\emph{Lemma 2.}
\[
\text{If }\rho_{v}\leq\rho_{v.sat}\text{,}\qquad\text{then }F>0\qquad
\text{for}\qquad\rho\in\left(  \rho_{v},\rho_{l}\right)  .
\]%
\[
\text{If }\rho_{v}\geq\rho_{v.spi}\text{,}\qquad\text{then }F<0\qquad
\text{for}\qquad\rho\in\left(  \rho_{v},\rho_{l}\right)  .
\]
The conclusions of Lemma 2 are illustrated in Fig. \ref{fig7}b. One can also
see that, if $\rho_{v.sat}<\rho_{v}<\rho_{v.spi}$, then $F(\rho,\rho_{v})$
does change its sign between $\rho=\rho_{v}$ and $\rho=\rho_{l}$.\medskip

\emph{Lemma 3.} Boundary-value problem (\ref{3.1}), (\ref{2.19}), (\ref{2.21})
admits a solution $\rho(r)$ only if $F(\rho,\rho_{v})$ changes sign within the
following open interval:%
\[
\min\left\{  \rho(r)\right\}  <\rho<\max\left\{  \rho(r)\right\}  .
\]
\emph{Proof.} Multiply equation (\ref{3.1}) by $r^{4}\mathrm{d}\rho
/\mathrm{d}r$, integrate from $r=0$ to $r=\infty$, and use (\ref{A.7}) to
rearrange the result in the form%
\[
\int_{0}^{\infty}r^{4}\frac{\mathrm{d}F}{\mathrm{d}r}\mathrm{d}r=\left[
\left(  r^{2}\frac{\mathrm{d}\rho}{\mathrm{d}r}\right)  ^{2}\right]
_{r\rightarrow\infty}-\left[  \left(  r^{2}\frac{\mathrm{d}\rho}{\mathrm{d}%
r}\right)  ^{2}\right]  _{r\rightarrow0},
\]
where the convergence of the integral on the left-hand side follows from the
large-$r$ asymptotics (\ref{A.2}). Integrating by parts and using boundary
conditions (\ref{2.19}), (\ref{2.21}), and (\ref{A.8}), one obtains%
\[
-4\int_{0}^{\infty}r^{3}F\,\mathrm{d}r=0.
\]
This identity proves the desired result.\medskip

Summarizing Lemmas 1--3, one can formulate the following theorem.\medskip

\emph{Theorem:} If $\rho_{v}\leq\rho_{v.sat}$ or $\rho_{v}\geq\rho_{v.spi}$,
boundary-value problem (\ref{3.1}), (\ref{2.19}), (\ref{2.21}) does not have solutions.

\section{Numerical solution of boundary-value problem (\ref{3.1}),
(\ref{2.19}), (\ref{2.21})\label{Appendix B}}

The boundary-value problem at issue involves two singular point, at $r=0$ and
$r=\infty$. Both pose difficulties when solving the problem numerically.

The singular point $r=0$ can be dealt with by observing that equation
(\ref{3.1}) and boundary condition (\ref{2.19}) imply%
\[
\rho\sim\rho_{0}+\frac{G(\rho_{0})-G(\rho_{v})}{3}r^{2}+\mathcal{O}%
(r^{4})\qquad\text{as}\qquad r\rightarrow0,
\]
where $\rho_{0}$ is to be determined together with the rest of the solution.
This asymptotics can be used for moving the boundary condition from $r=0$ to
$r=r_{1}$,%
\begin{equation}
\rho=\rho_{0}+\frac{G(\rho_{0})-G(\rho_{v})}{3}r_{1}^{2}\qquad\text{at}\qquad
r=r_{1}, \label{B.1}%
\end{equation}
where $r_{1}$ is a small positive number. Since (\ref{B.1}) involves an
undetermined parameter $\rho_{0}$, an extra boundary condition is required --
say,%
\begin{equation}
\frac{\mathrm{d}\rho}{\mathrm{d}r}=\frac{2\left[  G(\rho_{0})-G(\rho
_{v})\right]  }{3}r_{1}\qquad\text{at}\qquad r=r_{1}. \label{B.2}%
\end{equation}
The error associated with moving the boundary condition from $0$ to $r_{1}$
is, obviously, $\mathcal{O}(r_{1}^{4})$.

The singularity at $r=\infty$, in turn, was regularized using the
large-distance asymptotics (\ref{A.2}), which implies%
\begin{equation}
\frac{\mathrm{d}\rho}{\mathrm{d}r}=-\left(  \frac{1}{r_{2}}+\lambda\right)
\left(  \rho-\rho_{v}\right)  \qquad\text{at}\qquad r=r_{2}, \label{B.3}%
\end{equation}
where $r_{2}$ is a large positive number and $\lambda$ is given by
(\ref{A.3}). Given the exponential nature of the solution at large $r$, the
error associated with moving the boundary condition from $\infty$ to $r_{2}$
is exponentially small.

The regularized boundary-value problem comprising equation (\ref{3.1}) and
boundary conditions (\ref{B.1})--(\ref{B.3}) was solved for $\rho(z)$ and
$\rho_{0}$ using the MATLAB function BVP5c based on the three-stage Lobatto
IIIa formula \cite{KierzenkaShampine01}.

\bibliography{}

\end{document}